\documentclass[twocolumn,showpacs,nofootinbib,aps,10pt,prd]{revtex4}
\usepackage{graphicx,dcolumn,bm,amsfonts,amsmath,color,xcolor,amsthm,multirow,booktabs}
\usepackage[dvipdfm, colorlinks=true, citecolor=blue, linkcolor=blue,urlcolor=blue]{hyperref}
\usepackage{slashed,ulem,setspace}

\allowdisplaybreaks

\begin{document}
\title{Determining the kaon distribution amplitudes by fitting the $D\to K$ transition form factor to investigate the equivalence of two chiral LCSR frameworks}
\author{Fang-Ping Peng}
\author{Sheng-Bo Wu}
\author{Hai-Bing Fu}
\email{fuhb@gzmu.edu.cn}
\address{Department of Physics, Guizhou Minzu University, Guiyang 550025, P.R.China}

\begin{abstract}
In this work, we fit the $D\to K$ vector transition form factor (TFF) to the BESIII experimental data using a Monte Carlo (MC) method. We also calculate the TFF using QCD light-cone sum rules (LCSRs) with left-and right-handed chiral currents. For the TFF derived using the right-handed chiral current, the key input is the kaon twist-2 distribution amplitude (DA), for which we consider three models: the power-law parametrization ($\mathrm{R1}$), the truncated Gegenbauer expansion ($\mathrm{R2}$), and the light-cone harmonic oscillator (LCHO) model ($\mathrm{R3}$). The parameters of these models are determined by fitting the corresponding TFF predictions to the MC-fit results. For the TFF derived using the left-handed chiral current, the dominant input is the kaon twist-3 DA ($\mathrm{L4}$), which is constructed within the LCHO framework. Its parameters are determined from theoretical constraints rather than from the fit. In the large-recoil region, the resulting TFF are $f^{DK}_+(0)|_{\rm{R1}}=0.7381^{+0.0180}_{-0.0194}$, $f^{DK}_+(0)|_{\rm{R2}}=0.7320^{+0.0178}_{-0.0192}$, $f^{DK}_+(0)|_{\rm{R3}}=0.7324^{+0.0180}_{-0.0191}$, and $f^{DK}_+(0)|_{\rm{L4}}=0.7343^{+0.0459}_{-0.0573}$.
We then calculate the decay widths and branching fractions of the semileptonic decays $D\to K e\nu_e$ and extract the CKM matrix element
$|V_{cs}|$. Finally, we predict the two important ratios, $\mathcal{R}_{\pi^-/K^-}^{e}$ and $\mathcal{R}_{\pi^0/K^0}^{e}$. All the obtained results are consistent with existing experimental measurements and theoretical predictions.
\end{abstract}
\maketitle

\section{Introduction}
Semileptonic $D$ meson decays $(D \to P \ell^+ \nu_{\ell}, P = \pi, K)$ constitute important observables for testing the Standard Model (SM) in charm physics and offer an ideal platform for precision tests of electroweak predictions as well as in-depth studies of the strong interaction.
These processes proceed via the flavor-changing charged-current transitions $c \to d, s$ of the charm quark. At tree level, their transition amplitudes are governed by the second-row Cabibbo-Kobayashi-Maskawa (CKM) matrix elements $V_{cd}$ and $V_{cs}$, which provide an ideal avenue for the precise determination of these CKM matrix elements.
Among these channels, $D \to K \ell^+ \nu_{\ell}$, whose transition is governed by $V_{cs}$, stands as one of the most representative semileptonic charm decays. As the largest element in the second row of the CKM matrix, $V_{cs}$ is close to unity. Consequently, this decay channel is a Cabibbo-favored process with a sizable branching fraction. It thus serves as an important probe for extracting $V_{cs}$, testing lepton flavor universality (LFU), and searching for new-physics signatures.
However, achieving high-precision predictions for its branching fraction and differential decay rate requires a reliable determination of the transition form factor (TFF) entering the $D \to K$ hadronic matrix element. This quantity encodes the nonperturbative dynamics of hadronization. Its precise evaluation relies on nonperturbative QCD methods and requires cross-checks against experimental measurements.
Therefore, making precise theoretical predictions and performing tests of the observables related to semileptonic $D$ meson decays remain essential for precision tests of the Standard Model, the extraction of CKM matrix elements, and the search for new physics signals beyond the SM, while also providing valuable dynamical information for an in-depth understanding of the internal structure of hadrons.
\begin{table*}[t]
\footnotesize
\renewcommand{\arraystretch}{1.2}
\caption{The observables in $D^0\to K^-e^+\nu_e$ and $D^+\to\bar{K}^0e^+\nu_e$ decays from experiment and theory include branching fractions, TFF $f_+^{DK}(0)$, and CKM matrix elements $|V_{cs}|$.}
\label {Data from different experimental groups}
\begin{tabular}{l l l l l}
\hline
Experiments&$\mathcal{B}(D^0 \to K^- e^+ \nu_e)$&$\mathcal{B}(D^+ \to \bar {K}^0 e^+ \nu_e)$&$f_+(0)$&$|V_{cs}|$  \\
\hline
Belle'06~\cite{Belle:2006idb}&$3.45(7)(20)$&$-$&$0.695(7)(22)$&$-$
\\
BaBar'07~\cite{BaBar:2007zgf}&$3.522(27)(45)(65)$&$-$&$0.727(7)(5)(7)$&$-$
\\
CLEO'08~\cite{CLEO:2007ntr}&$3.56(3)(9)$&$8.53(13)(23)$&$-$&$1.015(10)(11)(106)$
\\
CLEO'09~\cite{CLEO:2009svp}&$3.50(3)(4)$&$8.83(10)(20)$&$0.739(7)(5)$&$0.985(9)(6)(103)$
\\
BESIII'15~\cite{BESIII:2015tql}&$3.505(14)(33)$&$-$&$0.7368(26)(36)$&$0.9601(33)(47)(239)$
\\
BESIII'19~\cite{BESIII:2018ccy}&$-$&$-$&$0.7327(39)(30)$&$0.955(5)(4)(24)$
\\
BESIII'24~\cite{BESIII:2024slx}&$3.521(9)(16)$&$8.864(39)(82)$&$0.7366(11)(13)$&$0.9623(15)(17)(40)$
\\
BESIII'26~\cite{BESIII:2026ydr}&$3.548(6)(17)$&$8.928(25)(50)$&$0.7383(7)(14)$&$0.9639(9)(18)(40)$
\\
\hline
Theories&$\mathcal{B}(D^0 \to K^- e^+ \nu_e)$&$\mathcal{B}(D^+ \to \bar {K}^0 e^+ \nu_e)$&$f_+(0)$&$|V_{cs}|$  \\
\hline
LQCD'05~\cite{FermilabLattice:2004ncd}&$-$&$-$&$0.73(3)(7)$&$0.969(39)(94)(24)$
\\
LQCD'21~\cite{Chakraborty:2021qav}&$-$&$-$&$0.7380(44)$  &$0.9663(53)(39)(19)(40)$
\\
LQCD'23-I~\cite{Parrott:2022rgu}&$-$&$-$&$0.7441(40)$&$-$
\\
LQCD'23-II~\cite{FermilabLattice:2022gku}&$-$&$-$&$0.7452(31)$&$0.9589(23)(40)(15)(05)(95)$
\\
CCQM'19~\cite{Ivanov:2019nqd}&$3.63$&$9.28$&$0.78(12)$&$-$
\\
SPCA'20~\cite{Yao:2020vef}&$3.83(28)$&$-$&$0.756(36)$&$-$
\\
CSMs'22~\cite{Yao:2021pdy}&$3.90(17)$&$-$&$0.796(9)$&$0.974(10)$
\\
LFQM'26~\cite{Yang:2025gfz}&$3.70^{+0.01+0.05+0.01}_{-0.01-0.05-0.02}$&$9.39^{+0.05+0.12+0.03}_{-0.04-0.12-0.05}$&$0.789^{+0.002}_{-0.001}$&$-$
\\
\hline
\end{tabular}
\end{table*}

From an experimental perspective, multiple experimental collaboration groups such as Belle'06~\cite{Belle:2006idb}, BaBar'07~\cite{BaBar:2007zgf}, CLEO~\cite{CLEO:2004arv, CLEO:2005cuk, CLEO:2005rxg, CLEO:2007ntr, CLEO:2009svp} and BESIII~\cite{BESIII:2015tql, BESIII:2018ccy, BESIII:2024slx, BESIII:2026ydr} collaborations have conducted systematic measurements and studies on a series of observable quantities related to this decay process. These observable quantities include TFF, decay widths, branching fractions, and CKM matrix elements $|V_{cs}|$, etc. To facilitate the systematic comparison and analysis of these experimental measurement results, this paper summarizes the experimental data measured by each collaboration group in Table~\ref{Data from different experimental groups}, for intuitive comparison. From this table, it can be seen that there are certain differences between different measurements within the same collaboration group and between independent measurements from different collaboration groups. This situation directly prompts us to conduct systematic theoretical research on this type of semileptonic weak decay process, aiming to reduce the deviation between theoretical predictions and experimental data. This constitutes one of the core research motivations of this paper.

From a theoretical perspective, the hadron matrix elements in the $D\to K$ semileptonic decay can be parameterized and described by the TFF $f_+^{DK}(q^2)$. It is worth noting that the study of this TFF usually requires the use of non-perturbative methods. Currently, various non-perturbative methods have been employed to investigation it, such as lattice QCD (LQCD)~\cite{FermilabLattice:2004ncd, Na:2010uf, Chakraborty:2021qav, Parrott:2022rgu, FermilabLattice:2022gku}, the covariant confining quark model (CCQM)~\cite{Soni:2018adu, Ivanov:2019nqd}, the symmetry-preserving continuum approach (SPCA)~\cite{Yao:2020vef}, the continuum Schwinger function methods (CSMs)~\cite{Yao:2021pdy}, the light-front quark model (LFQM)~\cite{Cheng:2017pcq, Yang:2025gfz}, light-cone sum rules (LCSR)~\cite{Khodjamirian:2000ds, Wu:2006rd, Fu:2020uzy, Lu:2021mri}, the Isgur-Wise Functions (IWF)~\cite{Rahmani:2022zjn} and the relativistic quark model (RQM)~\cite{Faustov:2019mqr}.
We also present predictions from a variety of distinct theoretical frameworks in Table~\ref{Data from different experimental groups}, revealing numerical discrepancies among different schemes. These discrepancies can be attributed to differing treatments of nonperturbative QCD inputs across approaches: the kaon distribution amplitude (DA) adopted in phenomenological frameworks employs distinct functional forms and moment parameters. Meanwhile, systematic differences exist across theoretical frameworks with respect to factorization schemes, higher-order radiative corrections, quark mass inputs, and the description of SU(3) flavor symmetry breaking effects.
Since each theoretical approach has its respective strengths, for instance the LQCD offers reliable and precise calculations of hadronic form factors, while the QCD LCSR method can reveal the behavior characteristics of the long-distance distribution amplitude of light mesons with different twist structures. Notably, these two methods are complementary in their applicability: LQCD yields precise results in the high $q^2$-region, whereas LCSR offers better predictions in the low to intermediate $q^2$-region. The organic integration of these approaches can, to a certain extent, more systematically and comprehensively elucidate the physical effects~\cite{Melikhov:2000yu}, and assist us in gaining a deeper understanding of the physical mechanism underlying the $D \to K e \nu_e$ semileptonic decay process.

The LCSR method is based on the operator product expansion within the light-cone region ($x^2 \rightsquigarrow 0$), and all nonperturbative dynamical effect parameters are parameterized as the amplitude of the  light-cone distribution amplitude (LCDA). This technique has been extensively employed in the study of numerous semileptonic decay processes~\cite{Duplancic:2008ix, Momeni:2019uag, Descotes-Genon:2019bud, Cheng:2019tgh, Momeni:2020zrb, Emmerich:2018rug, Momeni:2017vae, Shen:2016hyv, Bharucha:2015bzk, Sun:2011ssd, Wang:2010tz, Wang:2008bw, Beneke:2018wjp, Du:2003ja, Yang:2005bv, Singh:2018yvt, Fu:2018yin, Huang:2022xny}. In contrast to the conventional QCD sum rules, the key distinction of LCSR lies in its parameterization of non-perturbative dynamics. Instead of employing vacuum condensates of various dimensions, LCSR characterizes these effects through LCDAs. Thus, obtaining accurate TFF via the LCSR approach is of great importance for future systematic studies.
On the other hand, the  Monte Carlo (MC) method enables effective utilization of existing experimental $D\to K$ TFF data for relevant calculations: by performing a global fit to the TFF datasets provided by multiple experimental collaborations, one can further extract the decay width, branching fraction, and CKM matrix element $|V_{cs}|$ for the $D\to K e\nu_e$ semileptonic decay, thereby facilitating a systematic comparison between theoretical results and experimental measurements. The combination of these two methods enables both mutual cross-validation and a more robust bridge between theory and experiment.
Predictions for the $D\to K$ TFF from various theoretical frameworks differ from one another, and the steadily improving precision of experimental measurements imposes stricter demands on the accuracy of theoretical computations. Noticeable discrepancies still persist between experimental data and theoretical results for the same decay channel.
Since the precise extraction of the CKM matrix element $|V_{cs}|$ depends critically on a reliable $D\to K$ TFF, we carry out a systematic and comprehensive study of the $D\to K$ semileptonic decays to improve the consistency between theoretical predictions and experimental measurements. This constitutes one of the core motivations of the present work.

As one of the most crucial nonperturbative parameters, the DA of the kaon encapsulates long-range dynamical information at low energy scales and is vital for determining the behavior of the TFF. Therefore, a detailed investigation of the DA contributes to enhancing the precision of theoretical calculations for the TFF. From a theoretical perspective, the twist-2 DA of the kaon has been studied using QCD sum rules (QCDSR)~\cite{Ball:2003sc} and LQCD~\cite{RQCD:2019osh}. It is known that the twist-2 DA of the kaon can be expanded in terms of Gegenbauer polynomials, whose expansion coefficients, known as Gegenbauer moments, can be computed via methods such as QCDSR. In practice, to avoid potential spurious oscillations introduced by higher-order moments and to maintain model simplicity, only the first few Gegenbauer moments are typically retained, resulting in a truncated form (TF). For instance, the twist-2 DA for the $\pi$ and kaon, calculated using QCDSR up to the second moments, serves as an example~\cite{LatticeParton:2022zqc}.
Meanwhile, the DA of the kaon can also be described by other phenomenological models, such as the power-law parametrization (PLP) model and the light-cone harmonic oscillator (LCHO) model.
This can provide a new theoretical perspective for understanding the structure of mesons and help to deepen the understanding of their internal dynamics. Through cross-validation among various DA models, the validity and applicability of these models in the semileptonic decay process can be systematically investigated. This constitutes another core motivation of the present work.

The remainder of this paper is organized as follows. In Sec.~\ref{Sec:2}, we introduce the theoretical framework for the semileptonic decay process $D \to K e \nu_e$ within the SM and present the techniques for calculating relevant observables. We then compute the crucial TFF for $D \to K$ semileptonic decay using the QCD LCSR approach, and provide a concise introduction to the kaon twist-2 $\phi_{2;K}(x,\mu)$ and twist-3 $\phi_{3;K}^{p,\sigma}(x,\mu)$ LCDAs. In Sec.~\ref{Sec:3}, we present the SM predictions for the $D \to K e \nu_e$ semileptonic decays and perform a detailed numerical analysis, including the TFF, differential decay widths, branching fractions, predictions for the CKM matrix elements and ratios of branching fractions. Finally, a summary is provided in Sec.~\ref{Sec:4}.
\section{Theoretical framework}\label{Sec:2}
The $D \to K$ TFFs is defined in terms of the relevant hadronic matrix element
\begin{align}
&\langle K(p_K) | \bar {s} \gamma_\mu c | D(p_D) \rangle \nonumber\\
&= 2\, f_+^{DK}(q^2) p_{K_\mu} + [f_+^{DK}(q^2)+f_-^{DK}(q^2)] q_\mu,
\end{align}
where $q = (p_D - p_K)$ denotes the momentum transfer and $p_{K_\mu}$ is the kaon four-momentum.
$f_{\pm}^{DK}(q^2)$ denote the vector and scalar TFFs, respectively. Since the contribution from the scalar TFF is strongly suppressed in $D\to K$ semileptonic decays, we focus our detailed analysis solely on the vector TFF.

\begin{figure}[h]
\begin{center}
\includegraphics[width=0.48\textwidth]{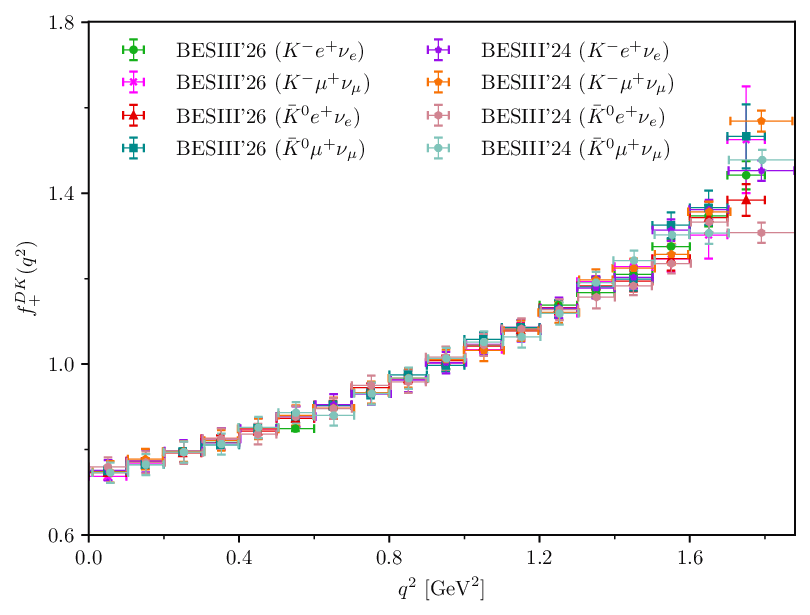}
\caption{The experimental data for the TFF $f_+^{DK}(q^2)$ from the BESIII~\cite{BESIII:2024slx, BESIII:2026ydr} collaborations.}
\label{Fig: MC and Experiment1}
\end{center}
\end{figure}

On the one hand, the BESIII Collaboration~\cite{BESIII:2024slx, BESIII:2026ydr} has presented the extensive experimental measurements of $D\to K$ vector TFF, as shown in Fig.~\ref{Fig: MC and Experiment1}. To fully exploit these experimental data, we adopt the MC method to extract the vector TFF $f_+^{DK}(q^2)$ via fitting, which is based on a model-independent formula derived solely from QCD dispersion relations as proposed by Boyd, Grinstein and Lebed (BGL)~\cite{Boyd:1994tt}.
\begin{align}
f_+^{DK}(q^2) = \frac{1}{P(q^2)\,\Phi(q^2)} \sum_{k=0}^{\infty} a_k (t_0) [z (q^2, t_0)]^k,
\label{Eq:TFF Fit}
\end{align}
here $a_k(t_0)$ are the real coefficients. The function $P(q^2) = z(q^2, m_{D_s^{*+}}^2)$ is the Blaschke factor that accounts for the pole of $f_+^{DK}(q^2)$ at $q^2 = m_{D_s^{*+}}$, where $z(q^2,t_0) = (\sqrt{t_+ - q^2}-\sqrt{t_+ -t_0})/(\sqrt{t_+ - q^2} + \sqrt{t_+ -t_0})$. The $\Phi(q^2)$  stands for outer functions, which are analytic in the unitary disc $|z|=1$ and are arbitrary, with no poles or branch cuts.
The choice of the factors $P(q^2)$ and $\Phi(q^2)$ must satisfy the constraint $\sum_{k=0}^{\infty} |a_k|^2 \leq 1$. This $z$-series expansion scheme is model-independent and naturally satisfies the requirements of analyticity and unitarity. The function $\Phi(q^2)$ is given by~\cite{BESIII:2024slx}:
\begin{align}
&\Phi(q^2) = \sqrt{\frac{1}{24 \pi \chi_V}} \Big(\frac{t_+ - q^2}{t_+ - t_0}\Big)^{\frac{1}{4}} \big(\sqrt{t_+ - q^2} + \sqrt{t_+}\big)^{-5} \nonumber\\
& \times \big(\sqrt{t_+ - q^2} + \sqrt{t_+ - t_0}\big) \big(\sqrt{t_+ - q^2} + \sqrt{t_+ - t_-}\big)^{\frac{3}{2}} \nonumber\\
& \times \big(t_+ - q^2 \big)^{\frac{3}{4}},
\label{Eq:Phiq2}
\end{align}
where $t_{\pm} = (m_D \pm m_K)^2$, $t_0 = t_+ \big(1 - \sqrt{1-t_-/t_+}\big)$ and the parameter $\chi_V=3/{32 \pi^2 m_c^2}$ is obtained from dispersion relations using perturbative QCD~\cite{Boyd:1995sq}. At this point, the complete formalism for the MC fit of the $D \to K$ TFF $f_+^{DK}(q^2)$ has been fully established.

On the other hand, within the LCSR framework, the TFF $f_+^{DK}(q^2)$ for this decay process can be calculated from the two-point correlation function:
\begin{align}
\Pi _{\mu}(p,q)&=i \int d^4x e^{iq\cdot x}\langle K(p)|T\{j_{\mu}(x), j_{D}^{\dagger}(0) \}| 0 \rangle \nonumber\\
& = F[q^2, (p+q)^2] p_\mu + \tilde F[q^2, (p+q)^2] q_\mu ,
\label{Correlation function}
\end{align}
where we employ two distinct chiral currents to compute the correlation functions. The first one is the right-handed chiral current, denoted as Scheme-R: $j^{\rm R}_{\mu}(x)=\bar d (x) \gamma_\mu  (1+\gamma_5) c(x)$ and $j^{\rm {\dagger R}}_{D}=\bar c(0) i (1+\gamma_5) d(0)$; the second is the left-handed chiral current, denoted as Scheme-L: $j^{\rm L}_{\mu}(x)=\bar d (x) \gamma_{\mu}  (1-\gamma_5) c(x)$ and $j^{\rm {\dagger L}}_{D}=\bar c(0) i (1-\gamma_5) d(0)$. The right-handed current correlator can effectively eliminate the uncertainties introduced by the twist-3 LCDAs and make the twist-2 LCDA the dominant contribution~\cite{Huang:2001xb}. This approach can effectively compensate for our insufficient understanding of the twist-3 wave functions and the lack of the corresponding $\mathcal{O}({\alpha_s})$ corrections.
The adoption of left-handed current correlator enables the isolation of expressions dependent exclusively on the twist-3 LCDAs, thereby circumventing the twist-2 contributions and subleading twist-4 effects and focusing the analysis on the physics at the twist-3 order~\cite{Zhong:2011jf}. This approach facilitates the investigation of the properties of twist-3 LCDAs and eliminates other irrelevant contaminating effects.

The correlation function is an analytic function of $q^2$ defined in both the timelike and spacelike $q^2$-regions. On the one hand, in the timelike $q^2$-region, one can insert a complete set of intermediate states with the same quantum numbers as the current $(\bar{c} i \gamma_5 d)$. After isolating the pole term of the $D$ meson to obtain the hadronic representation that can be decomposed into four Lorentz structures, and replacing the contributions from higher resonances and the continuum via a dispersion relation and quark-hadron duality, the hadronic representation of the correlation function is obtained. On the other hand, in the spacelike $q^2$-region, the correlation function can be calculated using the QCD OPE. Specifically, the OPE is performed near the light cone $x^2 \rightsquigarrow 0$ corresponding to the condition $(p + q)^2 - m_c^2 \ll 0$, and the momentum transfer $q^2$ is spacelike with $|q^2| \sim \mathcal{O}(1\,\text{GeV}^2)$ and $|q^2| \lesssim m_c^2$, which ensures the validity of the OPE. Finally, by employing the Borel transformation to suppress the contributions of higher-twist and continuum states, and matching the OPE result with the hadronic representation via quark-hadron duality, we derive the analytical expressions for the TFF within the framework of LCSR. The expressions are given in Appendix~\ref{Appendix-1}.

Based on the above discussion, we employ two independent approaches to perform calculations for the $D \to K e \nu_e$ semileptonic decay. The first approach combines experimental measurements and theoretical calculations: on the one hand, we fit the experimentally measured TFF data using the MC method. On the other hand, we derive the theoretical TFF expressions from the right-handed chiral current within the framework of QCD LCSR. Combining the experimental fit results with the theoretical calculations, we systematically compute and analyze the relevant physical observables. The second approach is entirely based on the left-handed chiral current and evaluates the TFF as well as the corresponding physical observables. A comparison of the results from the two approaches provides an effective test of the equivalence of the physical predictions obtained with the two chiral current schemes.

As the kaon DA is the most important non-perturbative input parameter in the theoretical calculation of TFF, its reasonable choice is the primary prerequisite for ensuring the reliability of the numerical TFF results. In the first approach, the twist-2 DA provides an important nonperturbative input for the TFF. To investigate the model dependence of the TFF on the DA, we employ three distinct DA models.

\noindent \textbf{~~The PLP model} is generalized from the symmetric Beta-form DA of the pion. Due to the mass difference between the strange and light quarks, the kaon exhibits a significant $\mathrm{SU}(3)_f$ flavor symmetry breaking. Consequently, its twist-2 DA no longer possesses the $x \leftrightarrow \bar{x}$ exchange symmetry. We generalize the PLP ansatz to an asymmetric Beta distribution involving two independent phenomenological parameters: $\phi_{2;K}^{\mathrm{PLP}}(x) = N x^{b_1} \bar{x}^{b_2}$,
where the parameters $b_1, b_2 > 0$ characterize the behavior of the distribution amplitude near the endpoints $x \to 0$ and $x \to 1$, respectively, and $N$ is the normalization constant. By substituting this expression into the normalization condition of the distribution amplitude and employing the Beta function identity $B(b_1+1, b_2+1) = [\Gamma(b_1+1)\Gamma(b_2+1)]/{\Gamma(b_1+b_2+2)}$, the normalization constant is determined. We thereby obtain the PLP model for the kaon twist-2 DA~\cite{RQCD:2019osh}:
\begin{align}
\phi_{2;K}^{\rm PLP}(x,\mu)|_{\rm R1} = x^{b_1} \bar{x}^{b_2} \frac{\Gamma(b_1+b_2+2)}{\Gamma(b_1+1)\Gamma(b_2+1)},
\label{PLP Tiwst-2}
\end{align}
here $\bar{x}=1-x$ and $b_{1,2}$ represent the model parameters.

\noindent \textbf{~~The TF model} is a widely adopted phenomenological model for the kaon twist-2 DA. It is constructed by truncating the Gegenbauer polynomial expansion at scale $\mu$ after the $N$th term. This expansion follows from solving the renormalization group equation~\cite{Lepage:1979zb, Efremov:1979qk}. In practical phenomenological analyses, only the first two Gegenbauer moments are retained since higher-order moments are numerically small and commonly neglected. It is expressed as
\begin{align}
\phi_{2;K}^{\rm TF}(x,\mu)|_{\rm R2} =6 x \bar x \Big[1 + c_1(\mu) C_1^{3/2}(\xi) + c_2(\mu) C_2^{3/2}(\xi) \Big],
\label{TF Tiwst-2}
\end{align}
where the coefficients $c_{1,2}(\mu)$ are the corresponding Gegenbauer moments, $C_{1,2}^{3/2}(\xi)$ denote the Gegenbauer polynomials and $\xi=2x-1$.

\noindent \textbf{~~The LCHO model} is a phenomenological approach based on the BHL framework~\cite{Wu:2007rt, Huang:2004su, Zhong:2011rg, Brodsky:1981jv, Cao:1997hw}. The valence-quark wave function in the rest frame is boosted to the light-cone form via the Wigner$\text{--}$Melosh transformation to obtain the light-cone wave function (LCWF) of the kaon. This LCWF factorizes into the Wigner$\text{--}$Melosh spin structure and the momentum-space harmonic-oscillator wave function~\cite{Fu:2014uea, Cheng:2017bzz}. After integrating out the transverse momentum $\mathbf{k}_{\perp}$, the twist-2 LCDA $\phi_{2;K}(x,\mu)$ for the kaon is obtained, which is expressed as follows:
\begin{align}
\phi _{2;K}^{\rm LCHO}&(x,\mu)|_{\rm R3} =\frac{\sqrt{3}A_{2;K}\beta _{2;K}\tilde{m}}{2\pi ^{3/2}f_K}
\nonumber\\
&\,\,  \times \varphi _{2;K}(x)\,\sqrt{x\bar{x}}\,\,\exp \left[ -\frac{\hat{m}_{q}^{2}x\,+\,\hat{m}_{s}^{2}\bar{x}-\tilde{m}^2}{8\beta _{2;K}^{2}x\bar{x}} \right]
\nonumber\\
&\,\,  \times \left\{ \mathrm{Erf}\left[ \sqrt{\frac{\tilde{m}^2+\mu ^2}{8\beta _{2;K}^{2}x\bar{x}}} \right] -\mathrm{Erf}\left[ \sqrt{\frac{\tilde{m}^2}{8\beta _{2;K}^{2}x\bar{x}}} \right] \right\} ,
\label{LCDA Tiwst-2}
\end{align}
where $\tilde{m} = \hat{m}_q x + \hat{m}_s \bar{x}$, $\hat{m}_q = 300$ MeV and $\hat{m}_s = 450$ MeV~\cite{Zhong:2022ecl}. The $\varphi_{2;K}(x)$ dominates the behavior of the WF's longitudinal distribution, which can be expressed as $\varphi_{2;K}(x) = (x \bar{x})^{\alpha_{2;K}} [1 + \hat{B}_{2;K} C_1^{3/2}(\xi)]$. It can be seen that the behavior of $\phi _{2;K}^{\rm R1}(x,\mu )$ in the LCHO model is determined by four parameters $A_{2;K}$, $\beta _{2;K}$, $\alpha_{2;K}$ and $\hat{B}_{2;K}$. It is worth noting that the three aforementioned twist-2 DA models contain undetermined parameters, which are determined via the reverse fit within the first approach.

Similarly, in the second approach, the twist-3 DA serves as an important nonperturbative input for the TFF. The kaon twist-3 DA based on the LCHO model takes the following form in the BHL framework~\cite{Zhong:2011rg}:
\begin{align}
\phi _{3;K}^{p,\sigma \rm}(x&,\mu)|_{\rm L4}=\frac{A_{3;K}^{p,\sigma}(\beta _{3;K}^{p,\sigma})^2}{2\pi ^2}\varphi _{3;K}^{p,\sigma}(x)\exp \left[ -\frac{1}{8(\beta _{3;K}^{p,\sigma})^2} \right.
\nonumber\\
&\,\,       \times \left. \left( \frac{\tilde{m}_{q}^{2}}{x}+\frac{\tilde{m}_{s}^{2}}{\bar{x}} \right) \right] \left[ 1-\exp \left( -\frac{\mu ^2}{8(\beta _{3;K}^{p,\sigma})^2x\bar{x}} \right) \right] ,
\label{LCDA Tiwst-3}
\end{align}
where $\tilde{m}_q \simeq 0.30$ GeV, $\tilde{m}_s \simeq 0.45$ GeV~\cite{Zhong:2011rg} and $\varphi_{3;K}^{p,\sigma}(x) = 1 + \hat{B}_{3;K}^{p,\sigma} C_1^{(1/2,3/2)}(\xi) + \hat{C}_{3;K}^{p,\sigma} C_2^{(1/2,3/2)}(\xi)$. The four unknown parameters in the twist-3 LCDA $\phi_{3;K}^{p,\sigma}(x,\mu)$ can be determined by the normalization condition of the WF, the average value of the transverse momentum $\left\langle \mathbf{k}_\perp^2 \right\rangle^{p,\sigma}_K$~\cite{Guo:1991eb} and the first two Gegenbauer moments of the LCDA. In the specific calculation process, the average value of the transverse momentum $(\left\langle \mathbf {k}_ \perp^2 \right\rangle^{p,\sigma}_K)^{1/2} \approx 0.350~\rm{GeV}$~\cite{Guo:1991eb} and the first two Gegenbauer moments $a_{1,p}^{3;K}$, $a_{2,p}^{3;K}$, $a_{1,\sigma}^{3;K}$ and $a_{2,\sigma}^{3;K}$ are taken from Ref.~\cite{Nam:2006mb}. For $\phi_{3;K}(x,\mu)$, the adopted values are $a_{1,p}^{3;K}(1~ \rm{GeV})=0.020$, $a_{2,p}^{3;K}(1~\rm{GeV})=0.137$, $a_{1,\sigma}^{3;K}(1~\rm{GeV})=-0.004$ and $a_{2,\sigma}^{3;K}(1~\rm{GeV})=-0.007$. Based on the above constraints, we can further determine the model parameters of twist-3 LCDA in the second approach, which lays a theoretical foundation for the subsequent calculation of observables.
Subsequently, from systematic calculations of the physical observables for this semileptonic decay within the two schemes, the applicability and reliability of different DA models can be examined, the dependence of the TFF on the selected DA models can be explored, and it can be tested whether the chiral currents employed in the two schemes are equivalent.

Finally, to calculate the corresponding decay branching fraction, the following formula can be used to define the decay width of the semileptonic decay $D\to K$:
\begin{align}
&\frac{d\Gamma (D\rightarrow K\ell \nu _{\ell})}{dq^2}=\frac{G_{F}^{2}|V_{cs}|^2}{192\pi^3m_{D}^{3}}(1-\frac{m_{\ell}^{2}}{q^2}) ^2\biggr[(1+\frac{m_{\ell}^{2}}{2q^2})
\nonumber\\
&\hspace{1.6cm} \times\lambda^{3/2}(m_{D}^{2},m_{K}^{2},q^2)|f_{+}^{DK}(q^2)|^2+\frac{3m_{\ell}^{2}}{2q^2}
\nonumber\\
&\hspace{1.6cm} \times\,( m_{D}^{2}-m_{K}^{2})^2\,\lambda^{1/2}(m_{D}^{2},m_{K}^{2}\,,q^2)
\nonumber\\
&\hspace{1.6cm} \times\,|f_{0}^{DK}(q^2)|^2 \biggr],
\label{Eq:dGamma}
\end{align}
where $G_F = 1.1664 \times 10^{-5}~\rm{GeV^{-2}}$ is the Fermi constant and $|V_{cs}|$ is CKM matrix element. $\lambda(m_D^2, m_K^2, q^2)$ is the phase space factor. In the lepton mass limit $m_\ell \to 0$, the decay width receives contributions exclusively from the vector TFF $f_+^{DK}(q^2)$.
\section{NUMERICAL ANALYSIS}\label{Sec:3}
For the subsequent numerical analysis, the $c$-quark mass $m_c = 1.27 \pm 0.02 \, {\rm GeV}$, the kaon mass $m_K = 493.677 \pm 0.013 \, {\rm MeV}$, the $D^0$ meson mass $m_{D^0}=1864.84\pm0.05 \, {\rm MeV}$, and the $D^+$ meson mass $m_{D^+} =1869.5\pm0.4 \, {\rm MeV}$ are adopted ~\cite{ParticleDataGroup:2024cfk}. The $D$ and kaon meson decay constants are taken as $f_D = 212.0\pm0.7\,{\rm MeV}$ and $f_K =155.7\pm0.3\,{\rm MeV}$~\cite{FlavourLatticeAveragingGroupFLAG:2024oxs}, respectively.
\begin{figure}[h]
\begin{center}
\includegraphics[width=0.45\textwidth]{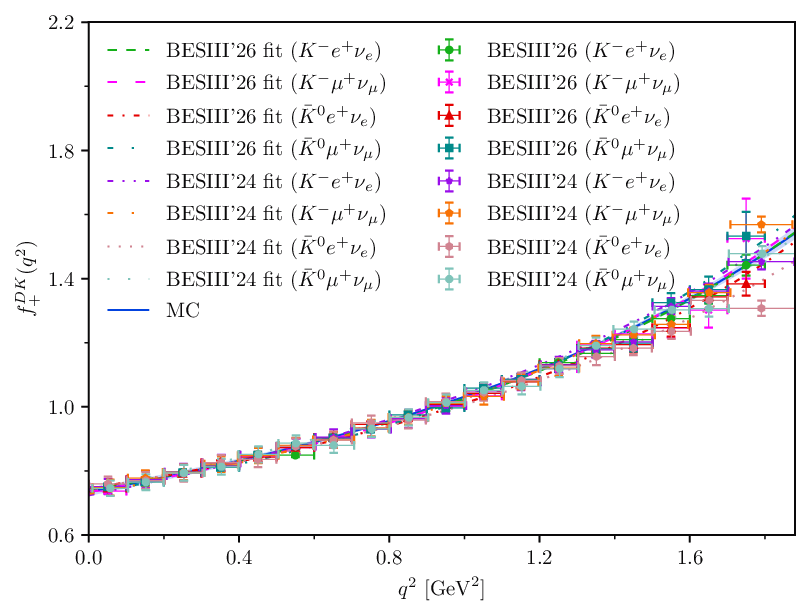}
\caption{The MC method was used to fit the results based on the experimental data points of the BESIII'24~\cite{BESIII:2024slx} and BESIII'26~\cite{BESIII:2026ydr} collaborations.}
\label{Fig: MC and Experiment2}
\end{center}
\end{figure}
We use the MC method to fit the TFF data points extracted by the BESIII Collaboration and ensure that these data points are evenly distributed throughout the study area, as shown in Fig.~\ref{Fig: MC and Experiment2}. It can be clearly seen that we extract eight sets of the TFF data points from the BESIII'24~\cite{BESIII:2024slx} and BESIII'26~\cite{BESIII:2026ydr} Collaborations, corresponding to the $D^0 \to K^- e^+ \nu_e$, $D^0 \to K^- \mu^+ \nu_{\mu}$, $D^+ \to \bar {K}^0 e^+ \nu_e$ and $D^+ \to \bar {K}^0 \mu^+ \nu_{\mu}$ decay channels, respectively.
At the same time, the MC method is used to fit the experimental data of each group. By combining the center values and their uncertainties of the eight experimental data sets, the total fitting results are further given. It is found that the overall fitting results show better behavior than the other eight individual experimental data sets, which not only reflects the reliability and rigor of the MC method but also helps to achieve the theoretical prediction accuracy of the physical observables of the semileptonic decay $D \to K$.
\begin{table*}[t]
\footnotesize
\renewcommand{\arraystretch}{1.2}
\caption{The BGL fitting method is applied to the experimental measurement of TFF data points, and the corresponding truncation parameter $a_k$, $k=(0,1,2,3)$ is successfully extracted. In this process, the BGL sequence was truncated at $N=3$.}
\label {Tab: truncation parameter}
\begin{tabular}{l l l l l l l}
\hline
&BESIII'24~\cite{BESIII:2024slx}~~~~~~~~&$a_0$~~~~~~~~&$a_1$~~~~~~~~&$a_2$
~~~~~~~~&$a_3$~~~~~~~~&$\chi^2/{\rm ndf}$ \\[1.0ex]
\hline
&$D^0 \to K^- e^+ \nu_e$
&$0.0318\pm0.00005$&$-0.0954\pm0.00053$
&$0.0645\pm0.05506$&$-0.1000\pm5\times10^{-6}$&$16.6639/14$
\\
&$D^0 \to K^- \mu^+ \nu_{\mu}$
&$0.0314\pm0.00004$&$-0.0900\pm4\times10^{-6}$
&$0.1868\pm0.04457$&$-0.0999\pm5\times10^{-6}$&$22.6572/14$
\\
&$D^+ \to \bar K^0 e^+ \nu_e$
&$0.0307\pm0.00003$&$-0.0629\pm4\times10^{-6}$
&$0.1286\pm0.07709$&$-0.0999\pm0.00001$&$20.1499/14$
\\
&$D^+ \to \bar K^0 \mu^+ \nu_{\mu}$
&$0.0317\pm0.00002$&$-0.0803\pm0.00006$
&$0.0466\pm0.01134$&$-0.1000\pm6\times10^{-6}$&$14.3906/14$
\\
\hline
&BESIII'26~\cite{BESIII:2026ydr}~~~~~~~~&$a_0$~~~~~~~~&$a_1$
~~~~~~~~&$a_2$~~~~~~~~&$a_3$~~~~~~~~&$\chi^2/{\rm ndf}$
\\
\hline
&$D^0 \to K^- e^+ \nu_e$
&$0.0313\pm0.00005$&$-0.0849\pm0.00003$
&$0.2583\pm0.23966$&$-0.1001\pm0.00006$&$17.7864/14$
\\
&$D^0 \to K^- \mu^+ \nu_{\mu}$
&$0.0313\pm0.00005$&$-0.0899\pm0.00007$
&$0.2789\pm0.01144$&$-0.1001\pm0.00004$&$17.7925/14$
\\
&$D^+ \to \bar K^0 e^+ \nu_e$
&$0.0309\pm0.00006$&$-0.0761\pm0.00004$
&$0.3422\pm0.15662$&$-0.1001\pm0.00005$&$17.2859/14$
\\
&$D^+ \to \bar K^0 \mu^+ \nu_{\mu}$
&$0.0313\pm0.00005$&$-0.1002\pm0.00006$
&$0.4635\pm0.25244$&$-0.1000\pm0.00004$&$18.7116/14$
\\
\hline
&MC&$0.0312\pm0.00009$&$-0.0855\pm0.00001$
&$0.1428\pm0.13339$&$-0.1001\pm0.00002$&$230.372/158$
\\
\hline
\end{tabular}
\end{table*}
\begin{figure}[h]
\begin{center}
\includegraphics[width=0.45\textwidth]{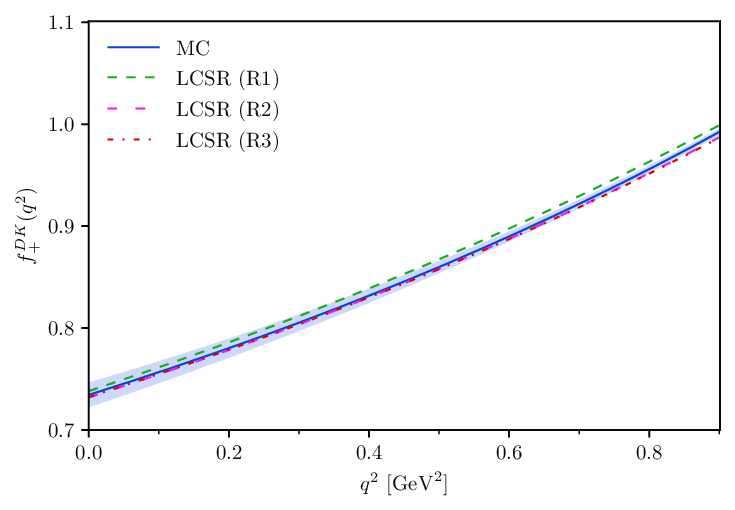}
\caption{The TFF $f_+^{DK}(q^2)$ versus $q^2$, where the solid line represents the MC fit result and the shaded bands indicate the corresponding uncertainty. The dashed line stands for the LCSR prediction.}
\label{Fig: TFF MC and LCSR}
\end{center}
\end{figure}

In order to further verify the reliability of the results obtained by the MC method, we use the fitted truncation parameters $a_k$, $k=(0,1,2,3)$ and the key indicator $\chi^2/\rm ndf$ to measure the fitting effect. These values are listed in Table~\ref{Tab: truncation parameter}. Among them, $\chi^2/\rm ndf$ is an important index to evaluate the fitting quality. When the value is closer to 1, the fitting results are more consistent with the experimental data. The data in Table~\ref{Tab: truncation parameter} show that the $\chi^2/\rm ndf$ values of each decay channel in the two experimental groups are close to 1. On the basis of comprehensive consideration of all data points of TFF, the $\chi^2/\rm ndf$ value of the fitting curve is also close to 1. The above results confirm the reliability of the results obtained based on the MC method.
\begin{table}
\begin{center}
\renewcommand{\arraystretch}{1.2}
\footnotesize
\caption{The central values of the DA parameters at $\phi_{2;K}(x,\mu_k)$ and $\phi^{p, \sigma}_{3;K}(x,\mu_k)$. Results for R1-R3 are obtained from the MC fit, and those for L4 are determined via parameter constraints.}
\label {Tab: Four model parameters}
\begin{tabular}{l l l l l l}
\cline{1-5}
\multirow{2}{*}{Twist-2~(R1)}~~~&$b_1$~~~~~~&~~~~~~&$b_2$~~~~~~& ~~~~~\\
 ~~~~~~      &0.9204      &      &1.7388 &
\\[2pt]
\hline
\multirow{2}{*}{Twist-2~(R2)}~~~&$c_1$~~~~~~&~~~~~~&$c_2$~~~~~~& ~~~~~\\
~~~~~~      &$-0.2562$      &      &0.0256 &
\\[2pt]
\hline
\multirow{2}{*}{Twist-2 (R3)}~~~~~~~~& $A_{2;K}$~~~~~~~~& $\beta_{2;K}$~~~~~~~~ & $\alpha_{2;K}$~~~~~~~~ & $\hat{B}_{2;K}$ \\
& 9.2614 & 0.7843 & $-0.4020$ & $-0.0343$
\\[2pt]
\hline
\multirow{4}{*}{Twist-3~(L4)}
~~~&$A_{3;K}^p$~~~~~~&$\hat{B}_{3;K}^{p}$
~~~~~~&$\hat{C}_{3;K}^{p}$~~~~~~&$\beta_{3;K}^p$~~~~~~\\
~~~~~~      &186.91      &0.4004      &$-1.1763$      &0.4915
\\
\cline{2-5}
~~~~~~&$A_{3;K}^{\sigma}$~~~~~~&$\hat{B}_{3;K}^{\sigma}$
~~~~~~&$\hat{C}_{3;K}^{\sigma}$~~~~~~&$\beta_{3;K}^{\sigma}$~~~~~~\\
~~~~~~      &299.28      &0.1114      &$-0.0231$      &0.3917
\\
\hline
\end{tabular}
\end{center}
\end{table}
We employ the BGL MC method to fit the $D \to K$ vector TFF $f_+^{DK}(q^2)$ in the range $0 \le q^2 \le (m_D - m_K)^2 \approx 1.88~\rm{GeV}^2$.
By combining this with the analytic TFF expression in the first approach, we determine the unknown parameters in the kaon twist-2 DA $\phi_{2;K}(x, \mu)$ for the R1, R2, and R3 models.
Meanwhile, for the twist-3 DA that acts as the nonperturbative input in the analytic TFF expression under the second approach, its unknown parameters can be well determined via the LCHO model together with four sets of constraint conditions.
After obtaining the numerical values of the above model parameters, we can acquire the complete behavior of the kaon twist-2 and twist-3 LCDA $\phi_{2;K}(x, \mu)$ and $\phi_{3;K}(x, \mu)$. On this basis, the vector TFF $f_K^{DK}(q^2)$ can be further computed within the framework of LCSR, which provides an essential prerequisite for the subsequent numerical evaluation of various physical observables.
Before executing the above strategy, the continuum threshold $s_0$ and the Borel parameter $M^2$ need to be determined first. Both of them can be determined according to the self-consistency criteria of the QCDSR~\cite{Tian:2023vbh}. Specifically, the continuum threshold $s_0$ should be close to the square of the mass of the first excited state of the $D$ meson. The selection of Borel window ensures contribution of continuous state is less than $30\%$, and contribution of high twist is kept as small as possible.
\begin{figure}[h]
\begin{center}
\includegraphics[width=0.45\textwidth]{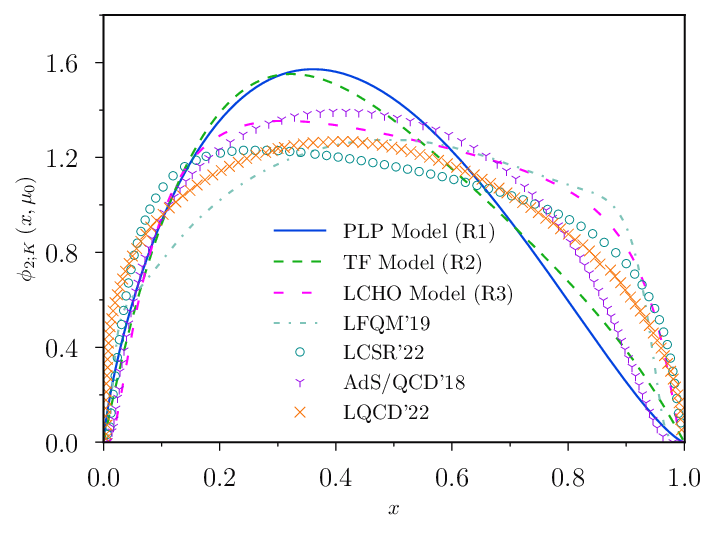}
\caption{The behavior of $\phi_{2;K}(x,\mu_0)$ for the three schemes at the scale $\mu_0=1.0~\rm {GeV}$. The  LFQM'19~\cite{Dhiman:2019ddr}, LCSR'22~\cite{Zhong:2022ecl}, AdS/QCD'18~\cite{Momeni:2017moz} and LQCD'22~\cite{LatticeParton:2022zqc} are also used for comparison.}
\label{Fig: twist-2 DA}
\end{center}
\end{figure}

\begin{figure}[h]
\begin{center}
\includegraphics[width=0.45\textwidth]{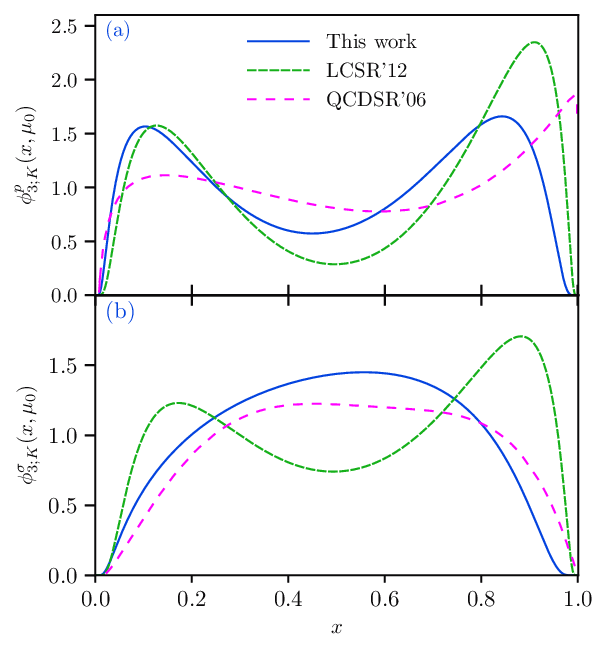}
\caption{Curves of the kaon twist-3 LCDAs $\phi_{3;K}^{p}(x,\mu_0)$ and $\phi_{3;K}^{\sigma}(x,\mu_0)$. The LCSR'12~\cite{Zhong:2011rg} and QCDSR'06~\cite{Ball:2006wn} are also used for comparison.}
\label{DAtwist-3}
\end{center}
\end{figure}

According to the above criteria, we determine the continuum threshold $s_0$ and the Borel parameter $M^2$ for the two methods, respectively. Specifically, for the first method, we adopt $s_0^{\rm R}=10.5\pm0.5\, \rm {GeV^2}$ and $M^2=32\pm1\, \rm {GeV^2}$; for the second approach, $s_0^{\rm L}=2.82\pm0.2\, \rm {GeV^2}$ and $M^2=15\pm0.5\, \rm {GeV^2}$ are used.
Given that LCSRs are only applicable in the low and intermediate $q^2$ regions, we further test the fit quality of the TFF $f_+^{DK}(q^2)$. Fig.~\ref{Fig: TFF MC and LCSR} presents the $q^2$ dependence of $f_+^{DK}(q^2)$ from the MC fit and three different LCSR schemes within $0 \le q^2 \le 0.9~\rm {GeV^2}$. In Fig.~\ref{Fig: TFF MC and LCSR}, the MC fitting results are basically coincident with the TFF under the three LCSR schemes.
After determining the continuum threshold $s_0$ and the Borel parameter $M^2$, the parameters of twist-2 LCDA $\phi_{2;K}(x,\mu_k)$ at the scale $\mu_k = \sqrt{m_D^2 - m_c^2} \simeq 1.4~\rm{GeV}$ are determined by using the vector TFF $f_+^{DK}$ obtained by MC fitting and combining with Eq.~\eqref{PLP Tiwst-2}-\eqref{LCDA Tiwst-2}.
Correspondingly, by combining four sets of parameter constraints, we can fully determine the four unknown parameters in the twist-3 kaon LCDA $\phi^{p,\sigma}_{3;K}$.
The results are listed in Table~\ref{Tab: Four model parameters}. Furthermore, we obtain the behavior of the twist-2, 3 LCDAs and compare it with other theoretical predictions, which we present as the corresponding curves in Fig.~\ref{Fig: twist-2 DA} and Fig.~\ref{DAtwist-3}.
For the purpose of comparison, the models in literature for kaon twist-2 LCDAs $\phi_{2;K}(x,\mu_0)$ such as LFQM'19~\cite{Dhiman:2019ddr}, LCSR'22~\cite{Zhong:2022ecl}, AdS/QCD'18~\cite{Momeni:2017moz} and LQCD~\cite{LatticeParton:2022zqc} are also shown in Fig.~\ref{Fig: twist-2 DA}.
It is observed that our predictions for the twist-2 LCDA under the three schemes exhibit a single-peak behavior, while R1 and R2 differ from R3. The results obtained from R1 and R2 are consistent with the theoretical predictions of AdS/QCD'18 in terms of trend, whereas the distribution profiles of R3 agree with the conclusions of LFQM'19, LCSR'22, and LQCD'22. The discrepancy in the behavior of $\phi_{2;K}(x,\mu_0)$ among R1, R2, and R3 may originate from the different theoretical models employed: the R1 scheme is based on the PLP model, the R2 scheme is constructed using the TF model, and the R3 scheme adopts the LCHO model.
The distribution curves of the kaon twist-3 LCDA $\phi^{p, \sigma}_{3;K}(x,\mu_0)$ are shown in Fig.~\ref{DAtwist-3}, where the predictions from LCSR'12~\cite{Zhong:2011rg} and QCDSR'06~\cite{Ball:2006wn} are also presented for comparison.
It can be observed that $\phi_{3;K}^{p}(x,\mu_0)$ is consistent with the result from LCSR'12. $\phi_{3;K}^{\sigma}(x,\mu_0)$ exhibits a smooth single-peak distribution. This is similar to the results of QCDSR'06, both of which are single-peak structures. It should be emphasized that the value of the parameter $\hat{C}^{\sigma}_{3;K}$ has a decisive influence on whether the distribution is a unimodal or bimodal structure. In contrast, the distribution of LCSR'12 shows a double-peak behavior with highs at both ends and lows in the middle.

Using these parameters, we compute the $D \to K$ TFF and list its value at the large recoil point in Table~\ref{Tab: TFFs}.
For comparison, we also present the predictions from BaBar'07~\cite{BaBar:2007zgf}, CLEO'09~\cite{CLEO:2009svp}, BESIII~\cite{BESIII:2026ydr, BESIII:2024slx, BESIII:2018ccy, BESIII:2015tql}, LQCD~\cite{Parrott:2022rgu, FermilabLattice:2022gku, Chakraborty:2021qav, FermilabLattice:2004ncd}, CCQM~\cite{Ivanov:2019nqd, Soni:2018adu} and SPCA'20~\cite{Yao:2020vef}. One can see that although the deviation between our results and the CCQM'18 predictions lies within the error bands, our values are in good agreement with the remaining experimental and theoretical results and show particularly good consistency with BESIII and LQCD.
Specifically, maximum deviations of only $0.189\sigma$ and $0.753\sigma$ are found for the predictions of our five schemes compared with the BESIII'26 as well as the LQCD'23-I results, respectively.
Besides, it can be seen that the TFF $f^{DK}_+(0)$ values for R1 to L4 are close to each other but exhibit certain differences, which mainly arise from the different DA models and theoretical approach adopted. These results demonstrate that the various DA models and theoretical calculation schemes adopted in this work have only a negligible overall impact on the TFF of this semileptonic decay process, and meanwhile verify that these different DA models and corresponding theoretical methods possess favorable applicability and computational reliability within this decay system.
\begin{table}[h]
\footnotesize
\renewcommand{\arraystretch}{1.2}
\begin{center}
\caption{$D\to K$ TFF at large point $f_{+}^{DK}(0)$ with uncertainties. Meanwhile experimental and theoretical predictions are also listed here as a comparison.}
\label{Tab: TFFs}
\begin{tabular}{l l l l}
\hline
&Experiments~~~~~~~~~~~~~~~~~~~~~~&$f_+^{DK}(0)$ \\
\hline
&BaBar'07~\cite{BaBar:2007zgf}&$0.727(7)(5)(7)$
\\
&CLEO'09~\cite{CLEO:2009svp}&$0.739(7)(5)$
\\
&BESIII'15~\cite{BESIII:2015tql}&$0.7368(26)(36)$
\\
&BESIII'19~\cite{BESIII:2018ccy}&$0.7327(39)(30)$
\\
&BESIII'24~\cite{BESIII:2024slx}&$0.7366(11)(13)$
\\
&BESIII'26~\cite{BESIII:2026ydr}&$0.7355(7)(14)$
\\
\hline
&Theories~~~~~~~~~~~~~~~~~~~~~~&$f_+^{DK}(0)$ \\
\hline
&This work~(R1)&$0.7381^{+0.0180}_{-0.0194}$
\\
&This work~(R2)&$0.7320^{+0.0178}_{-0.0192}$
\\
&This work~(R3)&$0.7324^{+0.0180}_{-0.0191}$
\\
&This work~(L4)&$0.7343^{+0.0459}_{-0.0573}$
\\
&This work~(MC)&$0.7345^{+0.0121}_{-0.0121}$
\\
&LQCD'05~\cite{FermilabLattice:2004ncd}&$0.73(3)(7)$
\\
&LQCD'21~\cite{Chakraborty:2021qav}&$0.7380(44)$
\\
&LQCD'23-I~\cite{Parrott:2022rgu}&$0.7441(40)$
\\
&LQCD'23-II~\cite{FermilabLattice:2022gku}&$0.7452(31)$
\\
&CCQM'18~\cite{Soni:2018adu}&$0.78$
\\
&CCQM'19~\cite{Ivanov:2019nqd}&$0.78(12)$
\\
&SPCA'20~\cite{Yao:2020vef}&$0.756(36)$
\\
\hline
\end{tabular}
\end{center}
\end{table}

\begin{figure}[h]
\begin{center}
\includegraphics[width=0.45\textwidth]{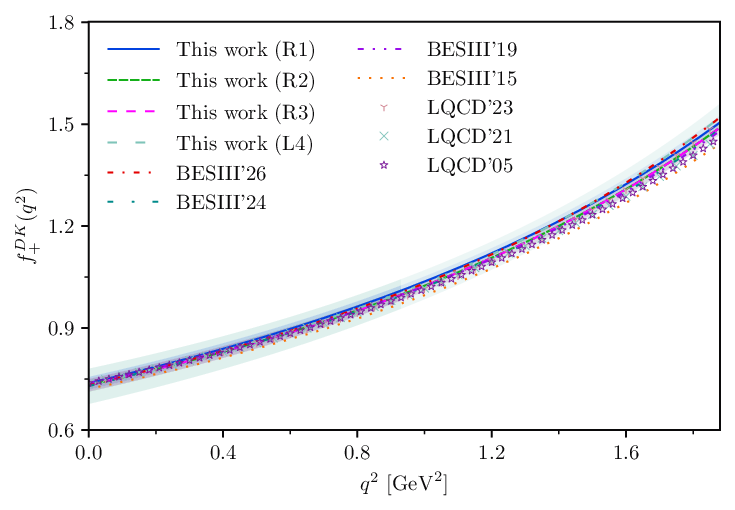}
\caption{The extrapolated TFF $f_+^{DK}(q^2)$ versus $q^2$. The solid lines and shaded bands indicate the central values and uncertainties of our predictions, respectively. In addition, we have also compared other theoretical predictions with experimental results.}
\label{Fig: TFF}
\end{center}
\end{figure}

The LCSR approach is valid only for low-to-intermediate squared momentum transfers $q^2\in[0,0.9\,{\rm GeV}^2]$, and its extrapolation to large $q^2$ becomes unreliable. To obtain the TFF over the physically allowed range $0 \leq q^2 \leq (m_D - m_K)^2 \approx 1.88\,{\rm GeV}^2$, we fit the numerical results using the simplified series expansion (SSE)~\cite{Bharucha:2010im}:
\begin{align}
f_{+}^{DK}(q^2) = \frac{1}{1 - q^2/m_{D_s^*}^2} \sum_{k=0,1,2} \beta_k \, z^k(q^2, t_0),
\end{align}
where $m_{D_s^{*+}}=2.1122~\rm{GeV}$ represents the pole mass. The fit results of free parameters $\beta_k$ are determined to make the quality of extrapolation $\Delta<1\%$. Then the quality of extrapolation $\Delta$ is defined as
\begin{equation}
\Delta = \frac{\sum_i |F_i(t) - F_i^{\text{fit}}(t)|}{\sum_i |F_i(t)|} \times 100,
\end{equation}
where $t\in[0,1/40,\cdots,1]\times 0.58\,{\rm GeV}$. After extrapolating the TFF across the whole physical $q^2$-region, the behavior of TFF $f_+^{DK}(q^2)$ can be determined, which is shown explicitly in Fig.~\ref{Fig: TFF}. Furthermore, the corresponding results calculated with BESIII~\cite{BESIII:2026ydr, BESIII:2024slx, BESIII:2018ccy, BESIII:2015tql} and the theoretical predictions from LQCD~\cite{Parrott:2022rgu, Chakraborty:2021qav, FermilabLattice:2004ncd} are also shown in Fig.~\ref{Fig: TFF}.
\begin{table}[h]
\begin{center}
\renewcommand{\arraystretch}{1.2}
\footnotesize
\caption{Fitted parameters $\beta_{1, 2}$ and extrapolation quality $\Delta$ for the TFF $f_+^{DK}(q^2)$ from the SSE method, including upper, central and lower limits.}
\label {Tab: SSETFFs}
\begin{tabular}{l l l l l}
\hline
&$f_{+}^{DK}(q^2)|_{\rm R1}$~~~~&$\beta_1$~~~~~~~~&$\beta_2$~~~~~~~~&$\Delta$ \\
\hline
&Upper&-1.2016&0.2270&$0.0065\%$
\\
&Center&-1.3871&-0.9671&$0.0170\%$
\\
&Lower&-1.5848&-2.1425&$0.0413\%$
\\
\hline
&$f_{+}^{DK}(q^2)|_{\rm R2}$~~~~&$\beta_1$~~~~~~~~&$\beta_2$~~~~~~~~&$\Delta$ \\
\hline
&Upper&-1.0936&1.0593&$0.0315\%$
\\
&Center&-1.2754&-0.1206&$0.0115\%$
\\
&Lower&-1.4694&-1.2828&$0.0107\%$
\\
\hline
&$f_{+}^{DK}(q^2)|_{\rm R3}$~~~~~~~~&$\beta_1$~~~~~~~~&$\beta_2$~~~~~~~~&$\Delta$ \\
\hline
&Upper &-1.0624&1.3120&$0.0319\%$
\\
&Center&-1.2450&0.1385&$0.0610\%$
\\
&Lower&-1.4292&-0.9382&$0.0912\%$
\\
\hline
&$f_{+}^{DK}(q^2)_{\rm L4}$~~~~&$\beta_1$~~~~~~~~&$\beta_2$~~~~~~~~&$\Delta$ \\
\hline
&Upper&-0.8465&3.3147&$0.0742\%$
\\
&Center&-1.2670&1.3799&$0.0855\%$
\\
&Lower&-1.7827&-0.4112&$0.0892\%$
\\
\hline
\end{tabular}
\end{center}
\end{table}
One can find that our predictions are compatible with the BESIII'26, BESIII'24, BESIII'15, LQCD'23 and LQCD'21 results over the whole $q^2$ range within uncertainties. In the low-$q^2$ region $(0 \le q^2 \le 0.9~\rm{GeV^2})$, with the exception of BESIII'19, they also show consistency with the other BESIII measurements and LQCD calculations. Overall, our predictions are in good agreement with the above experimental and LQCD results. Then, to check the rationality of the SSE extrapolation, the fitted parameters $\beta_{1;2}$ and the extrapolation quality $\Delta$ for $f_+^{DK}$, derived from the SSE method under the four schemes, are summarized in Table~\ref{Tab: SSETFFs}. One can see that the maximum value of $\Delta$ is only $0.0912\%$, which is far less than $1\%$, indicating good extrapolation effect.

The differential decay widths and branching fractions provide valuable information for understanding the internal structure of a particle and its interactions with other particles.
Employing the $D\to K$ TFF obtained via the LCSR approach, the Fermi constant, and the CKM matrix element $|V_{cs}|$~\cite{ParticleDataGroup:2024cfk}, we calculate the physical observables associated with the semileptonic decay $D \to K e \nu_e$.
Using Eq.~\eqref{Eq:dGamma}, we calculate the differential decay width of the semileptonic $D^0 \to K^- e^+ \nu_e$ decay in the whole $q^2$ region. Our results are presented as follows:
\begin{align}
&\Gamma(D^0 \to K^- e^+ \nu_e)|_{\rm R1} = 5.8646^{+0.2159}_{-0.2277}\times 10^{-14}\,{\rm GeV},
\nonumber\\
&\Gamma(D^0 \to K^- e^+ \nu_e)|_{\rm R2} = 5.7399^{+0.2116}_{-0.2232}\times 10^{-14}\,{\rm GeV},
\nonumber\\
&\Gamma(D^0 \to K^- e^+ \nu_e)|_{\rm R3} = 5.7383^{+0.2132}_{-0.2241}\times 10^{-14}\,{\rm GeV},
\nonumber\\
&\Gamma(D^0 \to K^- e^+ \nu_e)|_{\rm L4} = 5.6982^{+0.5482}_{-0.6389}\times 10^{-14}\,{\rm GeV},
\nonumber\\
&\Gamma(D^0 \to K^- e^+ \nu_e)|_{\rm MC} = 5.7976^{+0.8891}_{-0.8804}\times 10^{-14}\,{\rm GeV}.
\end{align}

\begin{figure}[h]
\begin{center}
\includegraphics[width=0.45\textwidth]{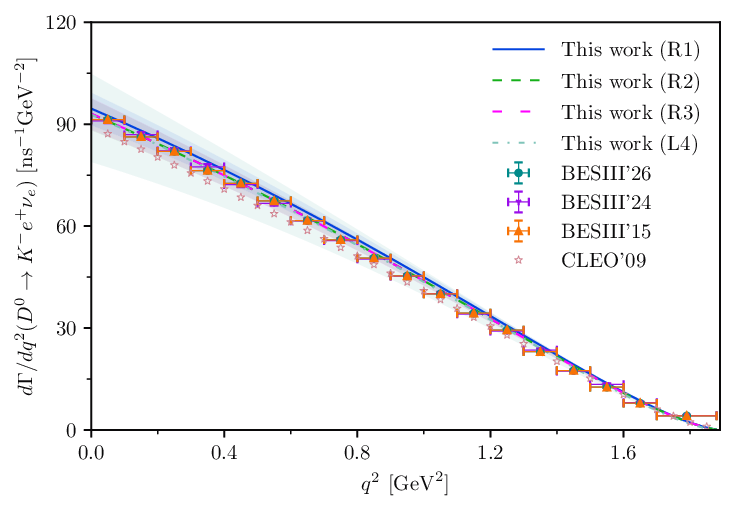}
\caption{The differential decay width of $D^0 \to K^- e^{+} \nu_{e}$ versus $q^2$. The shaded bands represent the theoretical uncertainties. Also shown for comparison are the experimental data from BESIII~\cite{BESIII:2024slx,BESIII:2015tql,BESIII:2026ydr} and CLEO'09~\cite{CLEO:2009svp}.}
\label{Fig: dGamma}
\end{center}
\end{figure}

The differential width results calculated by the five schemes are very close to each other. The relative differences among the central values of the decay widths obtained from the various schemes range from $0.03\%$ to $2.88\%$.
The resulting differential decay widths versus $q^2$ are shown in Fig.~\ref{Fig: dGamma}, along with the BESIII~\cite{BESIII:2024slx,BESIII:2015tql,BESIII:2026ydr} and CLEO'09~\cite{CLEO:2009svp} results for comparison.
It is evident from Fig.~\ref{Fig: dGamma} that the predictions of R2 and R3 almost coincide and are overall slightly lower than those of R1. This is mainly due to the relative differences in TFF: $0.015\sigma$ between R2 and R3, versus $0.216\sigma$ between R1 and R3, which consequently leads to the discrepancies among the three results. Overall, the predictions in this work are consistent with the BESIII and CLEO data within uncertainties. It is worth noting that our predictions fall between the BESIII and CLEO experimental data for $q^2 \in [1.7, q^2_{\rm{max}}]~\rm{GeV^2}$, which indicates that the discrepancy between the theoretical predictions and the experimental measurements still persists.

Furthermore, using the lifetimes of the $D^0$ and $D^+$ mesons $\tau_{D^0}=(0.410\pm 0.001)\,{\rm ps}$ and $\tau_{D^+}=(1.033\pm 0.005)\,{\rm ps}$~\cite{ParticleDataGroup:2024cfk}, we compute the branching fractions for the semileptonic decays $D^0 \to K^{-} e^{+} \nu_e$ and $D^+ \to \bar{K}^{0} e^{+} \nu_e$. Our results are compiled in Table~\ref{The branching fractions}. For comparison, the corresponding branching fractions obtained in PDG~\cite{ParticleDataGroup:2024cfk}, CLEO~\cite{CLEO:2005rxg,CLEO:2007ntr,CLEO:2009svp}, Belle~\cite{Belle:2006idb}, BESIII~\cite{BESIII:2024slx,BESIII:2026ydr}, LCSR~\cite{Wu:2006rd,Fu:2020uzy}, LFQM~\cite{Cheng:2017pcq, Yang:2025gfz}, CCQM'19~\cite{Ivanov:2019nqd}, SPCA'20~\cite{Yao:2020vef}, CSMs'21~\cite{Yao:2021pdy} and RQM'20~\cite{Faustov:2019mqr} are also presented in Table~\ref{The branching fractions}.
Our predictions exhibit deviations from some experimental and theoretical results, with the maximum deviation from the PDG values being about $0.988\sigma$ and from the BESIII data about $0.851\sigma$. It should be noted that the values for the $D^0$ channel for the MC scheme show deviations of $1.474\sigma$ and $0.945\sigma$ from the PDG and BESIII'26 data, respectively. These deviations may arise from uncertainties in the input parameters. Nevertheless, within the quoted uncertainties, our results remain consistent with most measurements.
In addition, the predictions from the five schemes are highly consistent with one another, with a maximum pairwise deviation of merely $0.3812\sigma$. On the one hand, this finding fully validates the reliability and applicability of the two theoretical approaches adopted in this work to the semileptonic decay channel under investigation, and it also demonstrates the equivalence of the two chiral currents for the decay process considered here. On the other hand, it clearly demonstrates that the three distinct DA parametrizations demonstrate strong robustness for the numerical predictions of this process.
\begin{table}[t]
\footnotesize
\renewcommand{\arraystretch}{1.2}
\caption{Our predictions for the branching fraction of semileptonic $D^0 \to K^{-} e^{+} \nu_e$ and $D^+ \to \bar{K}^{0} e^{+} \nu_e$ decays. For comparison, the experimental and theoretical results are also presented (in unit: $10^{-2}$).}. \label {Tab:dB}
\label{The branching fractions}
\begin{tabular}{l l l l}
\hline
&Experiments~&$\mathcal{B}(D^0 \to K^{-} e^{+} \nu_{e})$~~~&$\mathcal{B}(D^+\to \bar{K}^{0} e^{+} \nu_{e})$ \\
\hline
&CLEO'05~\cite{CLEO:2005rxg}     &$-$ &$8.71(38)(37)$
\\
&CLEO'08~\cite{CLEO:2007ntr}     &$3.56(3)(9)$ &$8.53(13)(23)$
\\
&CLEO'09~\cite{CLEO:2009svp}     &$3.50(3)(4)$ &$8.83(10)(20)$
\\
&Belle'06~\cite{Belle:2006idb}   &$3.45(10)(19)$ &$-$
\\
&BESIII'24~\cite{BESIII:2024slx} &$3.521(9)(16)$ &$8.864(39)(82)$
\\
&BESIII'26~\cite{BESIII:2026ydr} &$3.527(5)(16)$ &$8.918(25)(50)$
\\
\hline
&Theories~&$\mathcal{B}(D^0 \to K^{-} e^{+} \nu_{e})$~~~&$\mathcal{B}(D^+\to \bar{K}^{0} e^{+} \nu_{e})$ \\
\hline
&This work~(R1)   &$3.6530^{+0.1437}_{-0.1504}$ &$9.2039^{+0.3850}_{-0.4002}$
\\
&This work~(R2)   &$3.5754^{+0.1408}_{-0.1474}$ &$9.0082^{+0.3773}_{-0.3921}$
\\
&This work~(R3)   &$3.5743^{+0.1419}_{-0.1480}$ &$9.0056^{+0.3798}_{-0.3936}$
\\
&This work~(L4)   &$3.5520^{+0.0643}_{-0.0635}$ &$8.9427^{+0.9078}_{-1.0411}$
\\
&This work~(MC)   &$3.6113^{+0.3527}_{-0.4104}$ &$9.0987^{+0.1843}_{-0.1815}$
\\
&PDG~\cite{ParticleDataGroup:2024cfk} &$3.538\pm0.017$ &$8.81\pm0.07$
\\
&LCSR'06~\cite{Wu:2006rd}        &$3.20^{+0.47}_{-0.43}$ &$8.12^{+1.19}_{-1.08}$
\\
&LCSR'20~\cite{Fu:2020uzy}       &$3.370^{+0.176}_{-0.437}$ &$8.547^{+0.445}_{-1.108}$
\\
&LFQM'17~\cite{Cheng:2017pcq}    &$-$&$10.32(93)$
\\
&LFQM'26~\cite{Yang:2025gfz}     &$3.70(1)(5)(1)$ &$9.39(5)(12)(3)$
\\
&CCQM'19~\cite{Ivanov:2019nqd}   &$3.63$ &$9.28$
\\
&SPCA'20~\cite{Yao:2020vef}      &$3.834(282)$ &$-$
\\
&CSMs'21~\cite{Yao:2021pdy}      &$3.90(17)$ &$-$
\\
&RQM'20~\cite{Faustov:2019mqr}   &$3.56$ &$9.02$
\\
\hline
\end{tabular}
\end{table}

\begin{figure}[h]
\begin{center}
\includegraphics[width=0.45\textwidth]{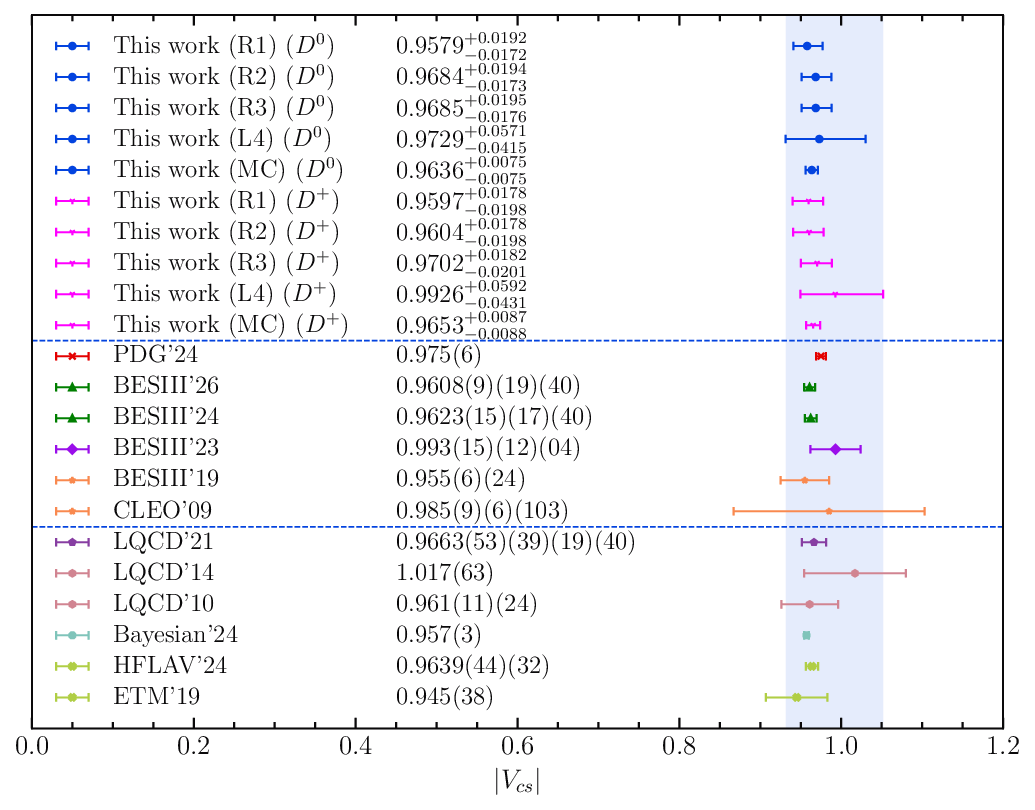}
\caption{Our prediction of $|V_{cs}|$ obtained from the decay channels $D^0 \to K^- e^+ \nu_e$ and $D^+ \to \bar{K}^0 e^+ \nu_e$, compared with other theoretical and experimental results of various channels.}
\label{Fig: Vcs}
\end{center}
\end{figure}

Employing the branching fractions from PDG average~\cite{ParticleDataGroup:2024cfk}, we extract the CKM matrix element $|V_{cs}|$ from the decays $D^0\to K^{-} e^+\nu_{e}$ and $D^+\to \bar{K}^{0} e^+\nu_{e}$. For comparison, we compile $|V_{cs}|$ results from various experimental measurements and theoretical predictions, as presented in Fig.~\ref{Fig: Vcs}.
These include results from the PDG~\cite{ParticleDataGroup:2024cfk}, BESIII~\cite{Zhang:2019tcs, BESIII:2023fhe, BESIII:2024slx, BESIII:2026ydr}, CLEO'09~\cite{CLEO:2009svp} Collaboration as well as LQCD~\cite{Chakraborty:2021qav, Donald:2013pea, Na:2010uf}, Bayesian'24~\cite{Bolognani:2024cmr}, HFLAV'24~\cite{HeavyFlavorAveragingGroupHFLAV:2024ctg} and ETM'19~\cite{Lubicz:2017syv} theory.
We observe that our five sets of predictions are compatible with most experimental and theoretical results within uncertainties, with the MC schemes and R1 showing optimal consistency with the recent BESIII'26 and Bayesian'24 data, respectively.
However, the central values from all five schemes are smaller than the PDG average and the BESIII'23 measurement. Some small central-value discrepancies appear locally: the R2 and R3 central values for the $D^0$ channel marginally exceed the BESIII'26 value, albeit with overlapping uncertainties, and all our results are lower by $0.87\% \text{--} 2.72\%$ compared with LQCD'22. In summary, our theoretical calculations are in good overall agreement with the reference data within the error range.

Finally, based on the branching fraction results from the PDG~\cite{ParticleDataGroup:2024cfk}: $\mathcal{B}(D^0 \to \pi^- e^+ \nu_{e})=(2.91\pm0.04) \times 10^{-3}$ and $\mathcal{B}(D^+ \to \pi^0 e^+ \nu_{e})=(3.72\pm0.17) \times 10^{-3}$, we calculate the ratios of branching fractions for each decay channel, which are summarized in Table~\ref{Tab:ratios}. The obtained $\mathcal{R}$ values are also compared with measurements from experimental collaborations, such as CLEO~\cite{CLEO:2005cuk,CLEO:2004arv,CLEO:2007ntr}. As shown in Table~\ref{Tab:ratios}, our calculated results agree well with the measurements from these experimental groups within errors and are consistent with the theoretical expectations of LFU. These ratios can be used to test the internal structure of $D$ mesons and the properties of related pseudoscalar mesons.
\begin{table}
\footnotesize
\renewcommand{\arraystretch}{1.2}
\caption{Ratios of branching fractions for exclusive semileptonic decays of $D^0$ and $D^+$. In order to compare, the results of different experimental groups are also given.}
\label{Tab:ratios}
\begin{tabular}{lll}
\hline
Methods&$\mathcal{R}_{{\pi^-}/{K^-}}^{e}$
&$\mathcal{R}_{{\pi^0}/{\bar K^0}}^{e}$ \\
\hline
This work~(R1)                &$0.0797^{+0.0012}_{-0.0010}$   &$0.0404^{+0.0019}_{-0.0018}$
\\
This work~(R2)                &$0.0814^{+0.0012}_{-0.0010}$   &$0.0413^{+0.0019}_{-0.0018}$
\\
This work~(R3)                &$0.0814^{+0.0012}_{-0.0010}$   &$0.0413^{+0.0019}_{-0.0018}$
\\
This work~(L4)                &$0.0819^{+0.0011}_{-0.0011}$   &$0.0416^{+0.0021}_{-0.0016}$
\\
This work~(MC)                &$0.0806^{+0.0015}_{-0.0006}$   &$0.0409^{+0.0019}_{-0.0019}$
\\
CLEO'05-I~\cite{CLEO:2004arv}  &$0.082(6)(5)$  &$-$
\\
CLEO'05-II~\cite{CLEO:2005cuk}   &$0.076(8)(2)$  &$-$
\\
CLEO'08~\cite{CLEO:2007ntr}     &$0.0841(32)(13)$ &$0.0437(27)(12)$
\\
\hline
\end{tabular}
\end{table}
\section{Summary}\label{Sec:4}
In this paper, we employ the MC method and the QCD LCSR approach to investigate the key physical quantity TFF in the semileptonic decay $D \to K$.
Firstly, we compile the experimental TFF data points reported by the BESIII Collaboration as shown in Fig.~\ref{Fig: MC and Experiment1}, and perform a fit to eight sets of independent experimental data points via the MC method to derive the TFF behavior, as shown in Fig.~\ref{Fig: MC and Experiment2}.
Taking all experimental data points into full account, we carry out a global fit to the TFF $f^{DK}_+(q^2)$ with the MC method according to Eq.~\eqref{Eq:TFF Fit}, and the resulting parameters are summarized in Table~\ref{Tab: truncation parameter}.
Furthermore, taking the $f^{DK}_+(q^2)$ computed via the first method as the fitting target, we determine the parameters of three twist-2 DA models for the kaon. Meanwhile, in the second method, the corresponding parameters of the kaon twist-3 LCDA are determined with the aid of four sets of constraint conditions, with the results listed in Table~\ref{Tab: Four model parameters}.
On this basis, we present the $f^{DK}_+(q^2)$ curves obtained from both the MC and LCSR schemes in the light-cone region, which are collectively displayed in Fig.~\ref{Fig: TFF MC and LCSR}.
Meanwhile, the specific behaviors of the kaon LCDA $\phi_{2;K}(x, \mu_0)$ and $\phi^{p, \sigma}_{3;K}(x, \mu_0)$ are presented in Fig.~\ref{Fig: twist-2 DA} and Fig.~\ref{DAtwist-3}, and compared with predictions from other existing models.

Then, based on the extracted DA parameters, we further derive the values of the TFF $f^{DK}_+(0)$ at the large recoil point. The corresponding results are summarized in Table~\ref{Tab: TFFs}, together with a systematic comparison against existing experimental measurements and theoretical predictions.
Subsequently, we extrapolate $f^{DK}_+(q^2)$ to the full physical $q^2$ region via the SSE approach. A comparison of the extrapolated results with other theoretical and experimental predictions is displayed in Fig.~\ref{Fig: TFF}.
To assess the validity of the extrapolation results, the expansion coefficients $\beta_k$ and the extrapolation quality $\Delta$ are presented in Table~\ref{Tab: SSETFFs}. It is found that $\Delta \ll 1\%$, which clearly demonstrates that the SSE extrapolation results are in excellent agreement with the LCSR calculations, fully validating the reliability of this extrapolation scheme.

Finally, we compute the differential decay widths and branching fractions for the two semileptonic decay channels $D^0 \to K^- e^+ \nu_e$ and $D^+ \to \bar K ^0 e^+ \nu_e$, and the corresponding numerical results are summarized in Fig.~\ref{Fig: dGamma} and Table~\ref{The branching fractions}. A systematic analysis of the tabulated data demonstrates that the branching fractions predicted in this work are consistent with existing experimental measurements within uncertainties. In addition, using the PDG averages of the branching fractions for $D^0 \to K^- e^+ \nu_e$ and $D^+ \to \bar{K}^0 e^+ \nu_e$, we extract the CKM matrix element $|V_{cs}|$, which is presented in Fig.~\ref{Fig: Vcs}. Using the branching fractions $\mathcal{B}(D^0 \to \pi^- e^+ \nu_e)$ and $\mathcal{B}(D^+ \to \pi^0 e^+ \nu_e)$ reported by the PDG, together with the $D\to K e\nu$ branching fractions predicted in this work, we derive the ratios $\mathcal{R}_{{\pi^-}/{K^-}}^{e}$ and $\mathcal{R}_{{\pi^0}/{\bar K^0}}^{e}$, which are listed in Table~\ref{Tab:ratios}. These calculation results are all in agreement with the current experimental measurements.

\acknowledgments
This work was supported in part by the National Natural Science Foundation of China under Grant No.12665017, No.12265010, the Project of Guizhou Provincial Department of Science and Technology under Grant No.MS[2025]219, No.CXTD[2025]030.

\appendix

\section{\NoCaseChange{The analytic expressions for the $D \to K$ vector TFF}}\label{Appendix-1}
Within the LCSR framework, the analytic expressions for the vector TFF of the $D \to K$ semileptonic decay from the left- and right-handed chiral currents are given as follows:
\begin{widetext}
\begin{align}
f_+^{\rm R}(q^2) &= \frac{m_c^2 f_K}{m_D^2 f_D}\, e^{\frac{m_D^2}{M^2}} \Bigg\{ \int_{\Delta}^1 du \exp\Bigg[- \frac{m_c^2 - (1 -u) \,(q^2 - u m_K^2)}{u M^2} \Bigg] \Bigg[\frac{\phi_{2;K}(u)}{u} - \frac{8 m_c^2 \,[g_1(u) + G_2(u)]}{u^3 M^4} + \frac{2g_2(u)}{u M^2} \Bigg] \nonumber\\
&  + \int_0^1 dv \int D\, \alpha_i \frac{\theta(\alpha_1 + v\, \alpha_3 - \Delta)}{(\alpha_1 + v \alpha_3)^2 M^2} \exp\Bigg[- \frac{m_c^2 - (1 - \alpha_i - v\, \alpha_3)\big[q^2 - (\alpha_1 + v\, \alpha_3) m_K^2\big]}{(\alpha_1 + v \alpha_3) M^2} \Bigg] \big[2 \Psi_{4;K}(\alpha_i) + 2 \nonumber\\
& \times \tilde{\Psi}_{4;K}(\alpha_i)  - \Phi_{4;K}(\alpha_i) - \tilde{\Phi}_{4;K}(\alpha_i)\big] \Bigg\},
\label{TFF right}
\\
f_+^{\rm L}(q^2) &=\mu_K \frac{m_c f_K}{m_D^2 f_K} e^{\frac{m_D^2}{M^2}} \Bigg\{ \int_{\Delta}^1 du\, \exp\Bigg[- \frac{m_c^2 - (1 -u) \,(q^2 - u\, m_K^2)}{u\, M^2} \Bigg] \Bigg[\mu\, \phi^p_{3;K}(u) + \frac{1}{6} \bigg( 2 + \frac{m_c^2 + q^2}{u\, M^2} \bigg) \phi^{\sigma}_{3;K}(u) \Bigg]
\nonumber\\
& - \frac{2 f_{3K}}{f_K \mu_K}\! \int^1_0 dv\! \int D \alpha_i \frac{\theta ( \alpha_1 + v \alpha_3 - \Delta)}{( \alpha_1 + v \alpha_3)^2} \exp\Bigg[\!- \frac{m_c^2 - q^2 (1 - \alpha_1 - v \alpha_3)}{(\alpha_1 + v \alpha_3) M^2}\! \Bigg] \Phi_{3;K}(\alpha_i) \Bigg[\! 1 - \frac{m_c^2 -q^2}{(\alpha_1 - v \alpha_3) M^2}\! \Bigg] \Bigg\},
\label{TFF left}
\end{align}
\end{widetext}
where $G_2(u) = \int_0^u g_2(v) dv$, $D \alpha_i = d \alpha_1 d\alpha_2 d \alpha_3 \delta ( 1 - \alpha_1 - \alpha_2 - \alpha_3 )$ and $\mu_K = m_K / ( m_u + m_s )$. The R and L denote the right- and left-handed chiralities, respectively. The integral lower limit $\Delta = \frac{1}{2 m_K^2} \Big[ \sqrt{(s_0 - q^2 - m_K^2)^2 + 4 m_K^2(m_c^2 - q^2)} - (s_0 - q^2 - m_K^2) \Big]$. Here, $g_i(u)$, $\Psi_{4;K}(\alpha_i)$, $\tilde \Psi_{4;K}(\alpha_i)$, $\Phi_{4;K}(\alpha_i)$, $\tilde \Phi_{4;K}(\alpha_i)$ stand for two-particle and three-particle twist-4 DAs, respectively.

\end{document}